\documentclass[journal=jacsat,manuscript=article]{achemso}

\usepackage[version=3]{mhchem} 

\newcommand{\affD}{SASTRA University, Tirumalaisamudram, Thanjavur, Tamilnadu-613401, India}

\author{Anirban Polley}
\email{anirban.polley@gmail.com}
\affiliation{\affD}

\title[An \textsf{achemso} demo]{Pure FENE Bond Potential for Soft Matter and Biological Simulations: Theory, HOOMD-blue Implementation, and Applications to Polymer, Colloidal, and Membrane Systems}
\abbreviations{FENE-WCA, HOOMD-Blue, MD}


\begin{document}


%
%
%


\begin{tocentry}
\centering
\includegraphics[width=8.25cm,height=3.5cm,keepaspectratio]{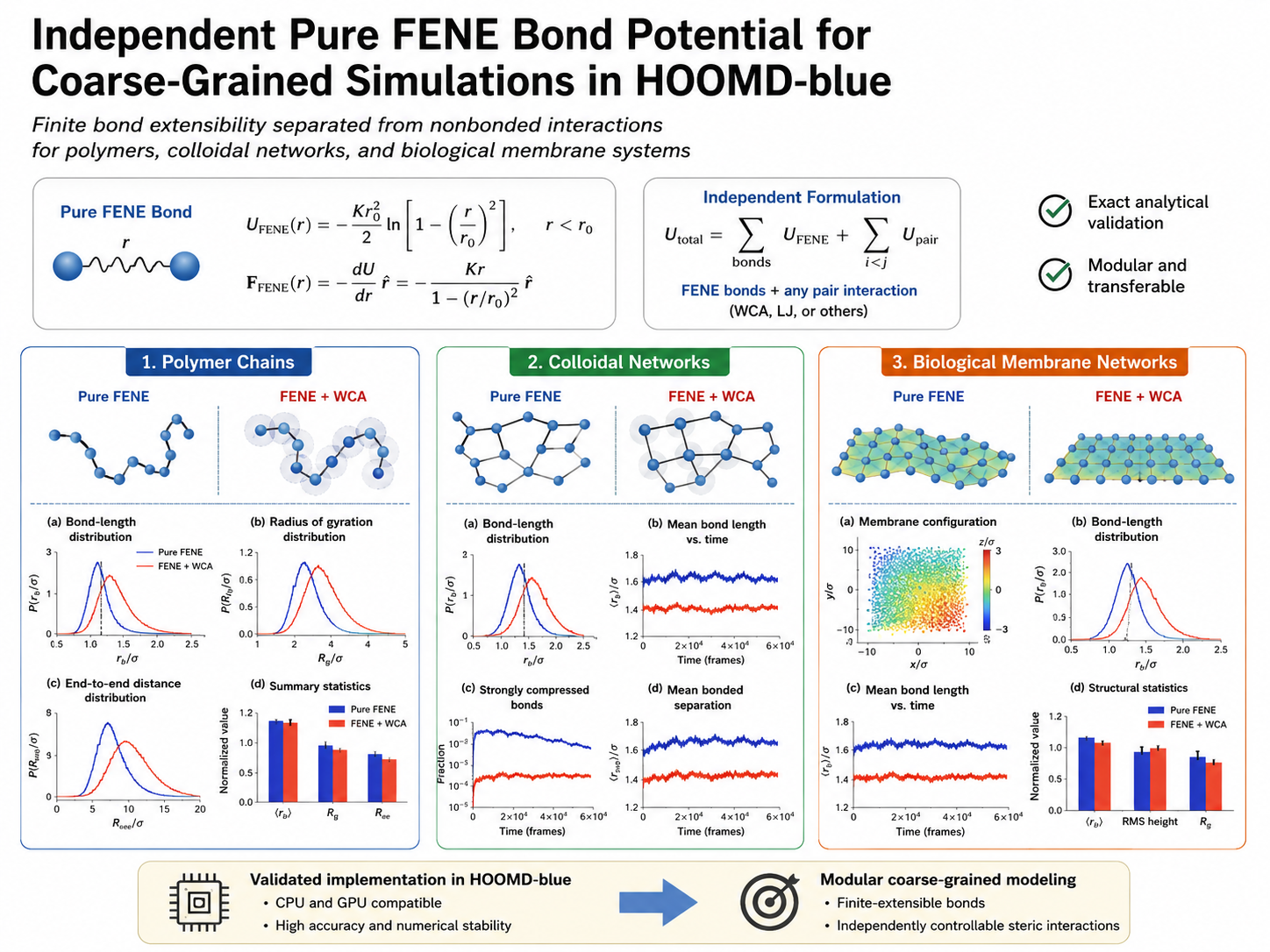}
\end{tocentry}
TOC Caption: Independent finite-extensible bond elasticity for
coarse-grained simulations of polymers, colloidal networks, and
biological membrane systems.

\begin{abstract}

The finitely extensible nonlinear elastic (FENE) potential is widely
used as a bonded interaction in coarse-grained simulations of
polymers, soft matter, colloids, and biological systems. In the
classical Kremer--Grest framework, FENE bonding is combined with a
short-range Weeks--Chandler--Andersen (WCA) interaction to provide
finite bond extensibility together with excluded-volume repulsion.
Although this combination is highly successful, it intrinsically
couples bonded elasticity to the nonbonded interaction, limiting the
ability to independently control these two contributions.
Here, we introduce a standalone FENE bond potential in HOOMD-blue in
which finite bond extensibility is implemented independently of the
choice of nonbonded interaction. This formulation allows the same
FENE bond potential to be combined with WCA, Lennard--Jones, or other
pair interactions without modifying the bonded interaction itself.
We derive the corresponding analytical energy and force expressions
and validate the HOOMD-blue implementation against exact
two-particle calculations, obtaining agreement to machine precision.
We further characterize the resulting bonded interactions through
their equilibrium bond distance and effective local stiffness,
demonstrating explicitly how the addition of nonbonded interactions
can alter these properties when they are not incorporated into the
bond potential.
We then demonstrate the utility of the standalone formulation in
coarse-grained polymer chains, colloidal networks, and mesh-based
biological membrane models. Across these systems, Pure FENE bonding
without short-range excluded-volume stabilization produces
pronounced structural contraction, whereas the addition of WCA
repulsion suppresses this collapse and preserves finite,
spatially extended structures. These results demonstrate that
separating finite bond extensibility from steric interactions
provides independent control over local bond mechanics and collective
structural organization. The standalone FENE formulation therefore
provides a modular framework for coarse-grained simulations in which
molecular connectivity and nonbonded interactions represent distinct
physical mechanisms.

\end{abstract}

\section{Introduction}

Coarse-grained molecular simulations have become an indispensable tool
for investigating the structural, dynamical, and mechanical properties
of polymers, colloidal assemblies, biomembranes, and other soft
condensed-matter systems. By reducing the number of degrees of freedom
while retaining the essential physical interactions, coarse-grained
models bridge the gap between atomistic simulations and continuum
descriptions, allowing access to experimentally relevant length and
time scales.~\cite{
Frenkel2002,
Noid2013,
Marrink2013,
Jin2022,
DAdamo2012,
Dhamankar2021,
Muller2006,
anirban_cell15,
anirban_prl16,
kamal2023}

One of the fundamental ingredients of a coarse-grained model is the
bonded interaction that maintains molecular connectivity. Among the
available bonded potentials, the finitely extensible nonlinear elastic
(FENE) potential has become particularly important in bead--spring
models as it imposes a finite upper bound on bond extension.
Unlike a harmonic spring, whose energy remains finite for any finite
bond length, the FENE potential diverges as the bond length approaches
a prescribed maximum extension. This property prevents arbitrarily
large bond stretching and provides a well-defined finite extensibility
for coarse-grained bonds.\cite{
Warner1972,
Bird1987,
Kremer1990,
Grest1986,
Larson1999,
Ottinger1996,
Morse2004}

The widespread adoption of the Kremer--Grest model established the
combination of FENE bonding and Weeks--Chandler--Andersen (WCA)
excluded-volume interactions as a standard coarse-grained description
of polymeric systems.~\cite{
Kremer1990,
Grest1986,
Weeks1971,
Kremer1989,
Auhl2003,
Likhtman2005,
Grest2007,
Padding2006}
The FENE term provides finite bond extensibility, while the WCA
interaction prevents unphysical overlap between particles. This
combination has been used extensively in studies of polymer melts,
solutions, rheology, glassy dynamics, entanglement, nanocomposites,
active polymers, and polymer networks.~\cite{
Doi1986,
Likhtman2002,
Graham2003,
Morse1999,
Larson1999,
Grest2007}

An important property of the FENE potential, however, is that finite
extensibility does not by itself imply a preferred finite equilibrium
bond length. For the standalone FENE interaction, the potential is
minimized at zero separation, while the divergence at the finite
extension $r_0$ limits the maximum bond length. A finite equilibrium
bond
length in the conventional Kremer--Grest model therefore emerges
from the competition between the attractive FENE bonding and the
short-range repulsive WCA interaction.~\cite{
Kremer1990,
Grest1986,
Likhtman2005,
Morse2004} This distinction is important
when the bonded and nonbonded contributions are considered as separate
physical ingredients. In particular, the addition of WCA shifts the
equilibrium bond distance away from zero and substantially changes the
local effective bond stiffness.~\cite{
Kremer1990,
Auhl2003,
Likhtman2005}

Beyond polymer physics, bonded interactions play a fundamental role
in many biological coarse-grained models. Triangulated and
mesh-based representations have been widely used to describe
deformable membranes, vesicles, and other soft interfaces, where
network connectivity can be separated from bending, self-avoidance,
and other mechanical contributions.\cite{
Gompper1994,
Gompper1995,
Gompper1997,
NoguchiGompper2006,
Cooke2005,
Muller2006,
Seifert1997,
Noguchi2009,
Fedosov2010,
Sadeghi2012}
Coarse-grained approaches have also been used to investigate the
structure and mechanics of biologically relevant molecular assemblies,
including full-length integrin activation.~\cite{
Bidone2019}
Cytoskeletal filaments, extracellular polymeric matrices, and other
cross-linked biological networks similarly require bonded interactions
that accurately describe stretching and force transmission.\cite{
MacKintosh1995,
Head2003a,
Head2003b,
Storm2005,
Broedersz2014,
Morse1998,
Heussinger2006,
Shankar2018,
Licup2015}

The conventional Kremer--Grest formulation provides a highly
successful reference model, but it is intrinsically associated with a
specific combination of FENE bonding and WCA excluded-volume
interactions.~\cite{
Kremer1990,
Grest1986,
Likhtman2005,
Auhl2003} Consequently, simulations based directly on this
formulation do not provide a straightforward way to retain the same
finite-extensible bonded interaction while systematically replacing
the nonbonded contribution with alternative interactions. Such
alternatives may include Lennard--Jones, Yukawa, Morse, electrostatic,
or experimentally derived pair potentials. Separating these two
contributions is particularly desirable when the bonded elasticity and
the nonbonded interaction are intended to represent different physical
mechanisms.

A harmonic bond combined with a separate pair potential provides one
possible route to such modularity. However, harmonic bonds have no
finite maximum extension and therefore allow arbitrarily large bond
lengths under sufficiently strong forces.~\cite{
milano2009,
Doi1986,
Frenkel2002} This limitation can become
important in systems undergoing large deformations, strong
interactions, nonequilibrium driving, or membrane remodeling. A
finite-extensible bonded interaction can instead impose an explicit
upper bound on bond extension while leaving the choice of nonbonded
interaction independent.

These considerations motivate a standalone implementation of the FENE
bond potential in which finite bond extensibility is represented solely
by the bonded interaction, while excluded-volume, attractive, or other
nonbonded interactions are specified independently. Such a formulation
retains the finite-extension characteristic of the FENE potential
without prescribing a particular pair interaction.~\cite{
Kremer1990,
Morse2004} It consequently
allows the same bonded interaction to be combined with WCA,
Lennard--Jones, or other pair potentials according to the physical
requirements of the system being modeled.

The distinction between bonded elasticity and nonbonded interactions is
also important for interpreting collective structural behavior. For
example, a system containing only attractive FENE bonds can undergo
pronounced contraction as the standalone FENE potential has no
finite-separation minimum. The addition of a short-range repulsive
interaction can counteract this contraction and stabilize a finite
structure. This provides a direct physical demonstration of why the
bonded and nonbonded contributions should be treated as independent
model components rather than as an inseparable interaction scheme.

In this work, we present the implementation of a standalone Pure FENE
bond potential within the molecular dynamics package HOOMD-blue.~\cite{
Anderson2020,
Glaser2015}
The implementation provides finite-extensible bonded interactions
independently of the choice of nonbonded pair potential. We first
derive the analytical energy and force expressions and compare the
Pure FENE interaction with harmonic, conventional FENE+WCA, and
harmonic+Lennard--Jones bond models. We determine their equilibrium
bond distances and effective local stiffnesses, highlighting the
distinct role of excluded-volume interactions in establishing a finite
equilibrium bond length. The numerical implementation is then
validated against exact two-particle calculations, demonstrating
agreement between the analytical and HOOMD-blue results to machine
precision.

Finally, we demonstrate the behavior of the standalone FENE
interaction in three representative coarse-grained systems: polymer
chains, colloidal networks, and a triangulated biomembrane. These
examples illustrate how the absence or presence of a separate
excluded-volume interaction influences collective structural
organization. In particular, Pure FENE produces pronounced structural
contraction in these systems, whereas the addition of WCA repulsion
suppresses collapse and preserves a finite, spatially extended
structure. The corresponding trajectories further illustrate the
distinct structural evolution produced by the two interaction
schemes. Together, these results establish the standalone Pure FENE
interaction as a modular framework for coarse-grained simulations in
which molecular connectivity and nonbonded interactions can be
controlled as distinct physical components.

\section{Theory}

Finite extensible nonlinear elastic (FENE) interactions constitute
one of the most widely used bonded potentials in coarse-grained
molecular simulations. Unlike harmonic springs, whose extension is
formally unbounded, the FENE potential imposes a finite maximum bond
length while providing a nonlinear restoring force. This property
prevents unphysical bond stretching and maintains molecular
connectivity, making the potential particularly suitable for
simulations of polymeric materials, soft colloids, and biological
macromolecules \cite{Bird1987,Kremer1990}.

\subsection{Finite Extensible Nonlinear Elastic (FENE) Potential}

The FENE interaction was originally introduced in the context of
finitely extensible polymer models and subsequently became central
to bead--spring molecular dynamics simulations.~\cite{
Warner1972,Bird1987,Kremer1990}

The FENE bond potential between two bonded particles separated by a
distance $r$ is given by

\begin{equation}
U_{\mathrm{FENE}}(r)
=
-\frac{1}{2}Kr_{0}^{2}
\ln\left(
1-\frac{r^{2}}{r_{0}^{2}}
\right),
\qquad
r<r_{0},
\label{eq:fene}
\end{equation}

where $K$ is the FENE spring constant, $r_0$ is the maximum allowed
bond extension, and $r$ is the instantaneous bond length.

The logarithmic divergence of Eq.~(\ref{eq:fene}) as
$r\rightarrow r_0$ ensures that the bond cannot reach or exceed the
maximum extension $r_0$. Thus, unlike a harmonic spring, the FENE
interaction provides an intrinsically finite bond-extension range.

It is important to distinguish finite extensibility from excluded
volume. The Pure FENE interaction constrains the bonded separation but
does not by itself introduce a nonbonded steric repulsion between
particles. Such excluded-volume effects must therefore be introduced
separately through an appropriate pair potential when required by the
physical model.

\subsection{Analytical Force}

The force associated with the FENE potential follows from

\begin{equation}
\mathbf{F}
=
-\nabla U_{\mathrm{FENE}}.
\end{equation}

For the radial potential in Eq.~(\ref{eq:fene}),

\begin{equation}
\frac{dU_{\mathrm{FENE}}}{dr}
=
\frac{Kr}
{1-r^{2}/r_{0}^{2}},
\end{equation}

and therefore the radial restoring force has magnitude

\begin{equation}
F(r)
=
-\frac{Kr}
{1-r^{2}/r_{0}^{2}}.
\label{eq:fene_force}
\end{equation}

Defining the interparticle displacement as

\begin{equation}
\mathbf{r}_{ij}
=
\mathbf{r}_{j}-\mathbf{r}_{i},
\end{equation}

with

\begin{equation}
r_{ij}=|\mathbf{r}_{ij}|,
\end{equation}

the force acting on particle $i$ is

\begin{equation}
\mathbf{F}_{i}
=
\frac{K}
{1-r_{ij}^{2}/r_{0}^{2}}
\mathbf{r}_{ij},
\label{eq:vector_force_i}
\end{equation}

while Newton's third law gives

\begin{equation}
\mathbf{F}_{j}
=
-\mathbf{F}_{i}.
\label{eq:vector_force_j}
\end{equation}

Equations~(\ref{eq:vector_force_i}) and
(\ref{eq:vector_force_j}) define the pairwise force convention used
for the implementation and numerical validation presented below.

\subsection{Properties of the FENE Potential}

The FENE interaction exhibits two limiting regimes that are
particularly important for coarse-grained simulations.

For small bond lengths,

\[
r\ll r_0,
\]

the logarithm in Eq.~(\ref{eq:fene}) can be expanded according to

\begin{equation}
\ln(1-x)
=
-x-\frac{x^2}{2}-\cdots,
\end{equation}

which gives

\begin{equation}
U_{\mathrm{FENE}}
\approx
\frac{1}{2}Kr^2
+
\mathcal{O}(r^4).
\end{equation}

Thus, the FENE interaction is harmonic to leading order at small
bond extension, with curvature determined by $K$.

In contrast, as

\[
r\rightarrow r_0^{-},
\]

the potential and force diverge,

\begin{equation}
U_{\mathrm{FENE}}
\rightarrow
+\infty,
\end{equation}

and

\begin{equation}
|F|
\rightarrow
+\infty.
\end{equation}

This divergence establishes a finite maximum bond extension and
strongly suppresses configurations approaching $r_0$.

An important consequence of the Pure FENE formulation is that the
potential itself has its minimum at

\begin{equation}
r_{\mathrm{eq}}=0.
\end{equation}

Thus, Pure FENE provides finite extensibility but does not impose a
finite preferred bond length. A finite equilibrium bond distance can
instead emerge when FENE is combined with a repulsive or attractive
nonbonded interaction.

\subsection{Comparison with Conventional Bond Potentials}

Several bonded interactions are commonly employed in coarse-grained
molecular simulations. Harmonic springs provide a simple linear
restoring force but permit arbitrarily large bond extensions. Morse
potentials, in contrast, possess a finite dissociation energy and can
therefore describe bond breaking. The classical Kremer--Grest model
combines FENE bonding with a purely repulsive Weeks--Chandler--Andersen
(WCA) interaction to provide finite bond extensibility together with
excluded-volume interactions \cite{Kremer1990,Weeks1971}.

The total interaction energy of such a model can be written as

\begin{equation}
U_{\mathrm{KG}}
=
\sum_{\mathrm{bonds}}
U_{\mathrm{FENE}}
+
\sum_{i<j}
U_{\mathrm{WCA}}.
\label{eq:kg}
\end{equation}

Although the FENE and WCA terms are mathematically distinct
contributions, conventional implementations often provide the
FENE--WCA combination as the standard bonded model. This can limit the
ability to systematically investigate finite bond extensibility while
independently changing the nonbonded interaction.

For example, alternative pair interactions may be required to
describe attractive, screened electrostatic, soft-core, or
experimentally derived interactions. An independent FENE bond
implementation allows such interactions to be introduced without
altering the bonded potential.

\subsection{Independent FENE Bond Formulation}

In the present implementation, the FENE interaction is introduced as
a standalone bonded potential,

\begin{equation}
U_{\mathrm{bond}}
=
U_{\mathrm{FENE}},
\label{eq:purefene}
\end{equation}

while nonbonded interactions are evaluated independently,

\begin{equation}
U_{\mathrm{total}}
=
\sum_{\mathrm{bonds}}
U_{\mathrm{FENE}}
+
\sum_{i<j}
U_{\mathrm{pair}}.
\label{eq:totalenergy}
\end{equation}

Here, $U_{\mathrm{pair}}$ may represent any compatible nonbonded
interaction available within the simulation framework, including
Lennard--Jones \cite{Jones1924}, WCA, Yukawa, Morse, Gaussian, electrostatic, or
user-defined pair potentials.

This separation provides independent control over molecular
connectivity and nonbonded interactions. Consequently, the effects of
finite bond extensibility can be investigated without simultaneously
fixing the functional form of the nonbonded interaction.

\subsection{Equilibrium Bond Length and Effective Stiffness}

The equilibrium bond distance is determined by the minimum of the
total interaction potential,

\begin{equation}
\left.
\frac{dU(r)}{dr}
\right|_{r=r_{\mathrm{eq}}}
=
0.
\label{eq:equilibrium}
\end{equation}

For Pure FENE alone, Eq.~(\ref{eq:equilibrium}) gives

\begin{equation}
r_{\mathrm{eq}}=0.
\end{equation}

Therefore, the Pure FENE interaction does not possess an intrinsic
finite preferred bond length.

When a nonbonded interaction is added, the equilibrium distance is
determined by the competition between the bonded and nonbonded
contributions. For the FENE+WCA model used here,

\begin{equation}
U_{\mathrm{FENE+WCA}}(r)
=
U_{\mathrm{FENE}}(r)
+
U_{\mathrm{WCA}}(r),
\end{equation}

which produces a finite equilibrium distance through the balance
between FENE attraction and short-range WCA repulsion.

The local effective bond stiffness is defined as

\begin{equation}
k_{\mathrm{eff}}
=
\left|
\frac{dF}{dr}
\right|_{r=r_{\mathrm{eq}}}.
\label{eq:keff}
\end{equation}

For the parameter set used in Fig.~\ref{fig2}, Pure FENE has
$r_{\mathrm{eq}}=0$ and a local curvature equal to the FENE spring
constant, $k_{\mathrm{eff}}=K$. Addition of WCA repulsion shifts the
equilibrium distance to approximately $r_{\mathrm{eq}}=0.961$ and
substantially increases the local curvature. The harmonic potential
has a fixed equilibrium distance of $r_{\mathrm{eq}}=1.0$ and
$k_{\mathrm{eff}}=K$, whereas addition of the Lennard--Jones
interaction shifts the equilibrium distance to approximately
$r_{\mathrm{eq}}=1.089$ and increases the local effective stiffness.

\subsection{Extension to Mesh-Based Biological Systems}

The mathematical form of the FENE interaction is independent of
molecular topology. Consequently, the same bonded potential can be
applied not only to linear polymers but also to triangulated networks
used to represent biological membranes, elastic capsules, and other
deformable soft materials.

For a triangulated membrane, the total bonded energy is

\begin{equation}
U_{\mathrm{mesh}}
=
\sum_{\langle ij\rangle}
U_{\mathrm{FENE}}(r_{ij}),
\end{equation}

where the summation extends over all connected edges of the
triangular mesh.

Additional membrane interactions, including bending rigidity, area
conservation, volume conservation, or membrane--protein interactions,
can be introduced independently of the FENE bond potential. This
modularity \cite{Noid2013,Jin2022} allows the bonded elasticity of the membrane mesh to be
controlled independently from its nonbonded interactions.

\subsection{Numerical Validation}

The analytical expressions given in Eqs.~(\ref{eq:fene}) and
(\ref{eq:vector_force_i}) provide exact benchmarks for validating the
numerical implementation.

In the present work, the computed bond forces and potential energies
are compared directly with the corresponding analytical expressions
over the physically accessible range

\[
0\leq r<r_0.
\]

The numerical results agree with the analytical expressions to machine
precision, confirming the correctness of the standalone FENE
implementation and establishing its equivalence to the prescribed
analytical potential.

\begin{figure*}[h!t]
\begin{center}
\includegraphics[width=16.0cm]{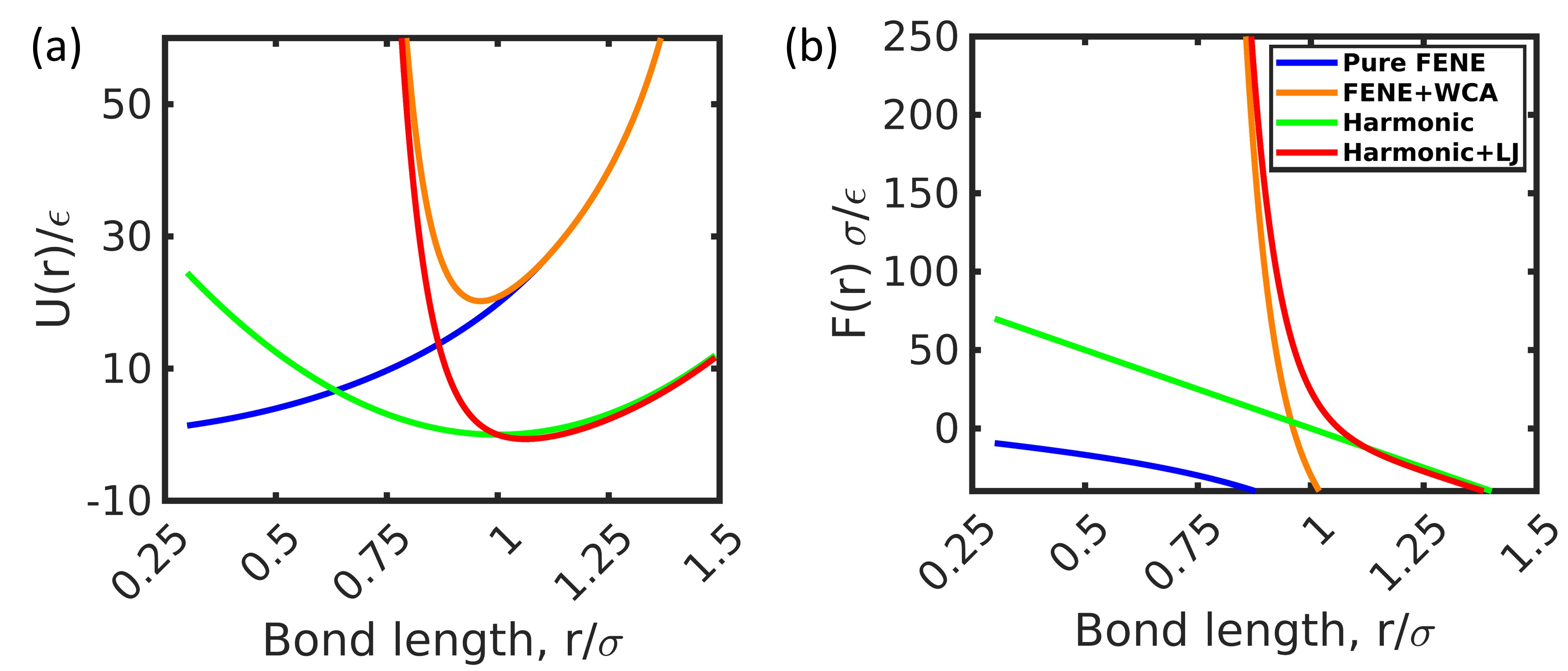}
\caption{
Analytical comparison of commonly used bonded and
bonded--nonbonded interaction models. (a) Potential energy and
(b) corresponding bond force as functions of bond length for
harmonic, Pure FENE, Kremer--Grest FENE+WCA, and
harmonic+Lennard--Jones interactions. Each potential is shifted so
that its minimum energy is zero. The Pure FENE interaction provides
finite bond extensibility determined solely by the FENE parameter
$r_0$, without introducing a preferred finite equilibrium bond
length. In contrast, coupling FENE or harmonic bonding with WCA or
Lennard--Jones interactions produces a finite equilibrium bond
distance through competition between bonded and nonbonded
interactions and modifies the local effective bond stiffness. The
comparison illustrates the distinction between finite bond
extensibility and the independent specification of nonbonded
interactions.
}
\label{fig1}
\end{center}
\end{figure*}

The equilibrium bond distance and local effective stiffness of the
four interaction models provide quantitative measures of how the
different bonded and nonbonded contributions modify local bond
mechanics.\cite{Helfrich1973,Lipowsky1991,EvansRawicz1990} For the Pure FENE potential, the equilibrium occurs at
$r_{\mathrm{eq}}=0$, with the local stiffness equal to the FENE spring
constant,
$k_{\mathrm{eff}}=K=30$. Thus, Pure FENE provides finite extensibility
but does not impose a preferred finite bond length. When the FENE bond
is combined with WCA repulsion, the competition between the FENE
bonding interaction and short-range steric repulsion produces a finite
equilibrium distance,
$r_{\mathrm{eq}}\approx0.961$, accompanied by a substantially larger
local effective stiffness,
$k_{\mathrm{eff}}\approx981$. For the harmonic bond,
$r_{\mathrm{eq}}=1.0$ and $k_{\mathrm{eff}}=K=30$, reflecting its
constant curvature. Addition of the Lennard--Jones interaction shifts
the equilibrium to $r_{\mathrm{eq}}\approx1.089$ and increases the
local effective stiffness to approximately $135$. These results
demonstrate that, once bonded and nonbonded interactions are coupled,
the equilibrium bond distance and local stiffness are properties of
the combined force-field construction rather than of the bonded
functional form alone.

\begin{figure*}[h!t]
\begin{center}
\includegraphics[width=16.0cm]{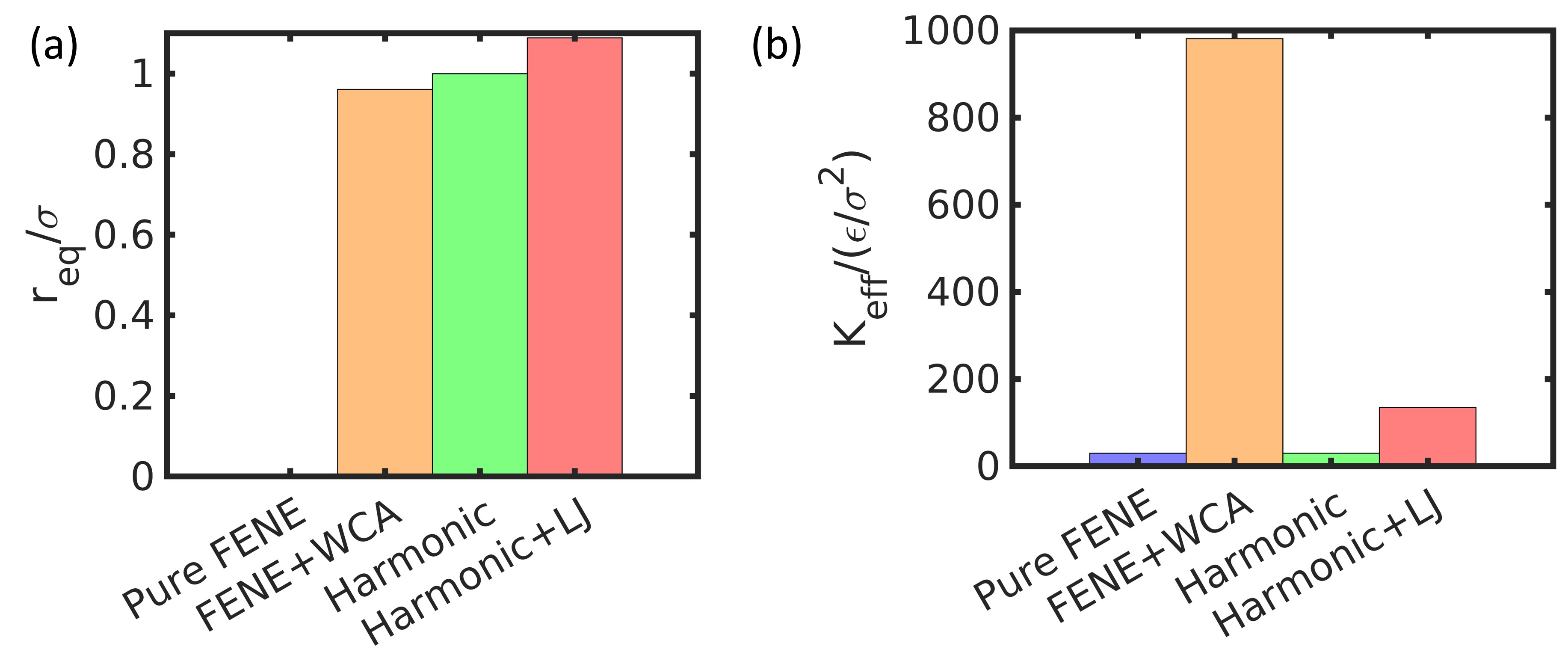}
\caption{
Comparison of the equilibrium bond distance and effective bond
stiffness for Pure FENE, FENE+WCA, harmonic, and
harmonic+Lennard--Jones interaction models.
(a) Equilibrium bond distance $r_{\mathrm{eq}}$ obtained from the
minimum of the corresponding total interaction potential.
(b) Effective bond stiffness
$k_{\mathrm{eff}}=
\left|\frac{dF}{dr}\right|_{r=r_{\mathrm{eq}}}$
evaluated at the equilibrium bond distance.
Pure FENE provides finite bond extensibility without an intrinsic
finite equilibrium bond length, giving $r_{\mathrm{eq}}=0$ for the
standalone potential. Addition of WCA excluded-volume repulsion
produces a finite equilibrium distance through competition between
FENE bonding and short-range steric repulsion. The harmonic and
harmonic+Lennard--Jones models possess intrinsic finite equilibrium
distances, with the Lennard--Jones contribution shifting the
equilibrium distance and increasing the local effective stiffness.
}
\label{fig2}
\end{center}
\end{figure*}

\begin{figure*}[h!t]
\begin{center}
\includegraphics[width=16.0cm]{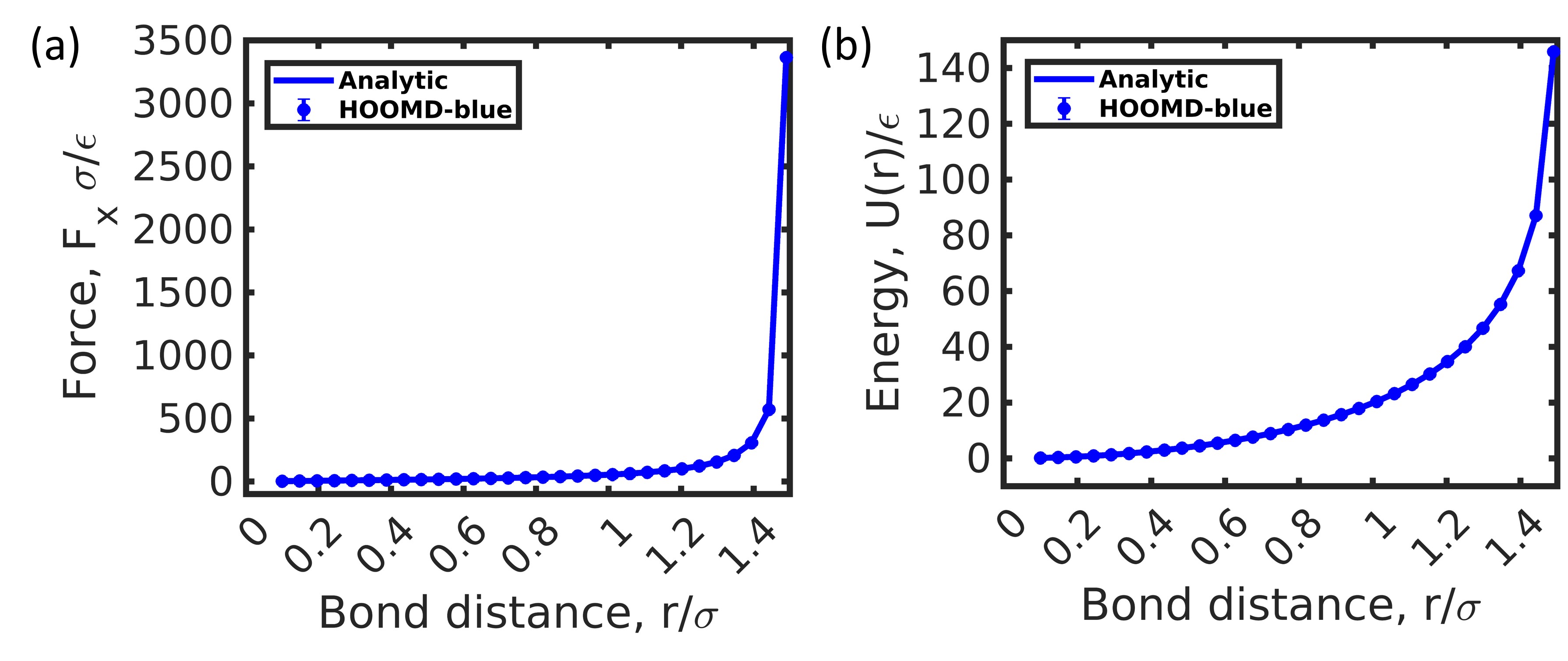}
\caption{
Validation of the HOOMD-blue pure FENE bond implementation using
analytical two-particle tests. (A) Force as a function of bond
distance, comparing HOOMD-blue results (symbols) with the analytical
pure FENE expression (solid line). (B) Potential energy as a function
of bond distance, comparing HOOMD-blue results (symbols) with the
analytical expression (solid line). The maximum relative errors are
$8.25 \times 10^{-15}$ for force and
$2.14 \times 10^{-15}$ for energy, demonstrating agreement to
machine precision.
}
\label{fig3}
\end{center}
\end{figure*}


\section{Applications}

The independent formulation of the FENE interaction is particularly
useful when the same finite-extensible bonded interaction must be
combined with different nonbonded interactions or applied to systems
with different molecular topologies. To demonstrate this versatility,
we consider three representative classes of coarse-grained systems:
linear polymer chains, cross-linked colloidal networks, and
mesh-based biological membrane models.~\cite{Gompper1994,Gompper1995,Gompper1997,NoguchiGompper2006} In all cases, the FENE
parameters are kept fixed, while the presence or absence of the
additional steric interaction allows the contribution of excluded
volume to be distinguished from that of bond elasticity.

\subsection{Polymer Chains}

We first examine the consequences of separating bond elasticity from
steric interactions in a coarse-grained polymer system. Linear
polymer chains containing 50 beads were simulated using either the
Pure FENE bond interaction or the conventional FENE+WCA combination.~\cite{
Kremer1990,Grest1986,Weeks1971,polley2013polymer}
The bonded parameters were fixed at $K=30$ and $r_0=1.5\,\sigma$,
while the WCA interaction was included only in the latter model.

\begin{figure*}[h!t]
\begin{center}
\includegraphics[width=16.0cm]{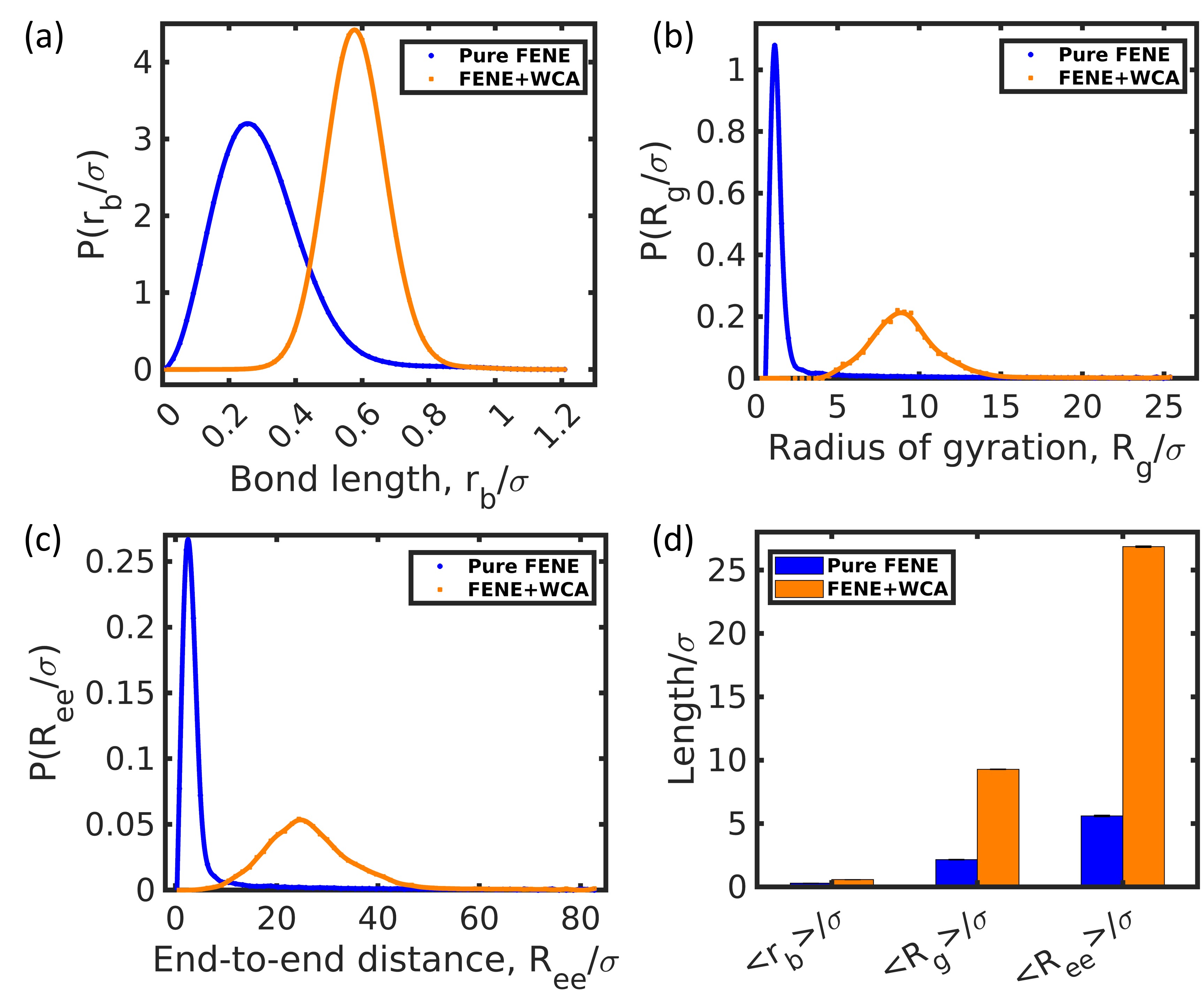}

\caption{
Polymer-chain statistics obtained using Pure FENE and FENE+WCA
interactions. (a) Bond-length distributions,
$P(r_b/\sigma)$. (b) Radius-of-gyration distributions,
$P(R_g/\sigma)$. (c) End-to-end distance distributions,
$P(R_{ee}/\sigma)$. (d) Mean bond length, radius of gyration, and
end-to-end distance for the two interaction models; error bars denote
the standard error of the mean. All lengths are reported in reduced
units of $\sigma$.
}
\label{fig4}
\end{center}
\end{figure*}

Figure~\ref{fig4} summarizes the resulting polymer statistics.
The bond-length distributions in Figure~\ref{fig4}a directly
illustrate the effect of the additional steric interaction on local
bond structure. As Pure FENE contains no short-range repulsive
contribution, its bond-length distribution reflects the intrinsic
statistics of the finite-extensible bond. Addition of the WCA
interaction modifies this distribution through the additional
excluded-volume constraint.

The conformational consequences are quantified through the radius of
gyration and end-to-end distance distributions shown in
Figure~\ref{fig4}b and c, respectively. These observables characterize
the global size and extension of the polymer chains and therefore
provide a direct measure of how local bonded and nonbonded interactions
propagate to chain-scale structure.

Figure~\ref{fig4}d compares the corresponding mean bond length,
radius of gyration, and end-to-end distance. The results demonstrate
that changing the nonbonded interaction while keeping the FENE bond
parameters fixed can modify both local bond statistics and global
polymer conformation.~\cite{Doi1986,Likhtman2002,
Graham2003,Morse1999} This illustrates the principal advantage of
the independent formulation: finite bond extensibility can be
preserved while the steric interaction is varied independently.

\subsection{Colloidal Networks}

We next consider a cross-linked colloidal network to examine the
effect of finite extensibility in a many-body connected structure.~\cite{Likos2001,
Zaccarelli2005,Sciortino2005}
The network contains $500$ particles connected by $750$ FENE bonds,
corresponding to an average coordination number of three. The same
bond parameters, $K=30$ and $r_0=1.5\,\sigma$, are used for both
models, while the FENE+WCA system additionally contains the
short-range WCA interaction.

\begin{figure*}[h!t]
\begin{center}
\includegraphics[width=16.0cm]{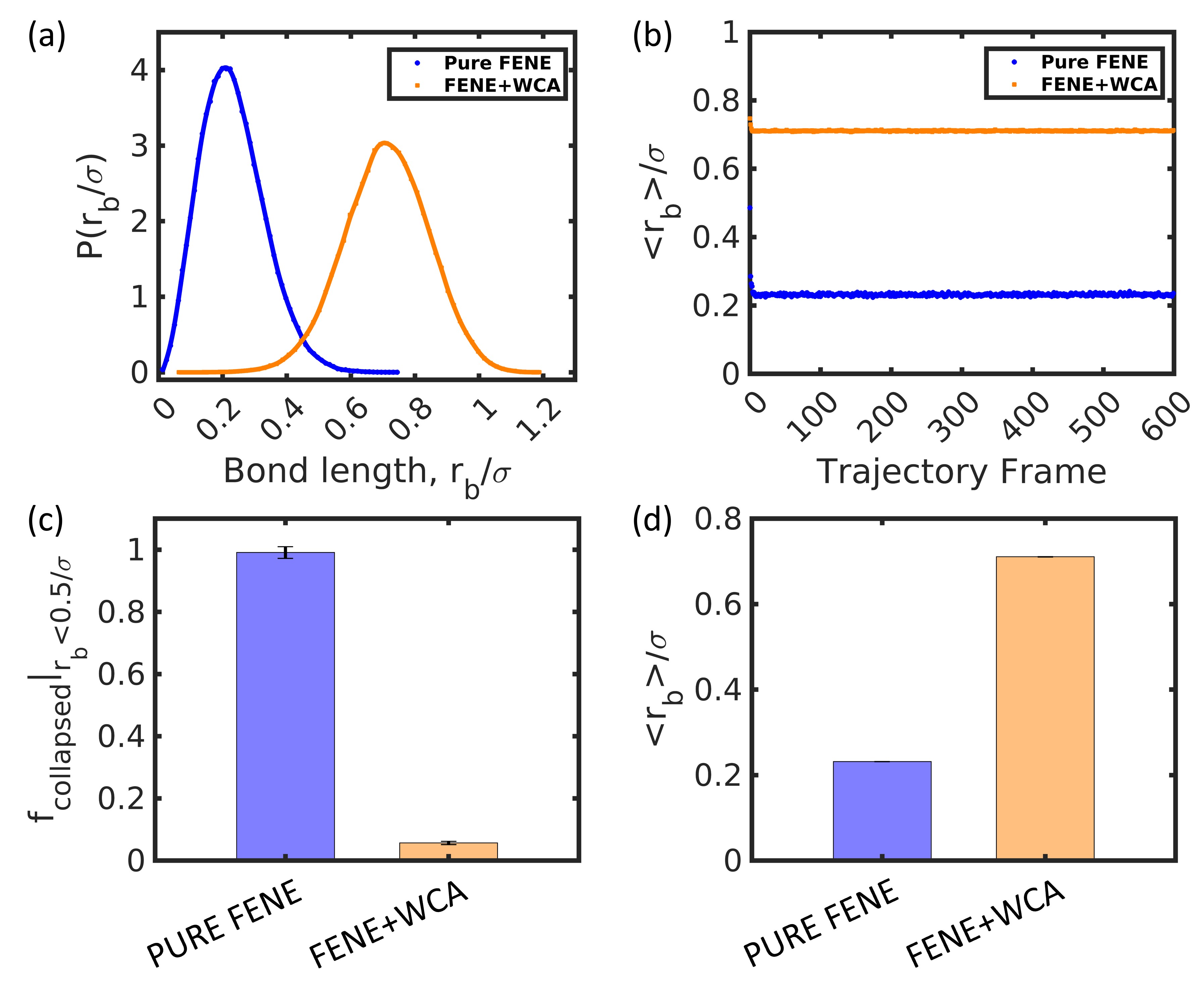}

\caption{
Colloidal-network structure and bonding statistics for Pure FENE and
FENE+WCA interactions. (a) Bond-length distributions. (b) Evolution
of the mean bond length during the simulation. (c) Fraction of
strongly compressed bonds. (d) Mean bonded separation. The comparison
demonstrates how the inclusion of an independent steric interaction
modifies local network structure while leaving the finite-extensible
bond interaction unchanged. All distances are reported in reduced
units of the particle diameter, $\sigma$.
}
\label{fig5}
\end{center}
\end{figure*}

The results are presented in Figure~\ref{fig5}. The bond-length
distribution in Figure~\ref{fig5}a demonstrates how the steric
interaction modifies the local bonded environment. Figure~\ref{fig5}b
shows the evolution of the mean bond length during the simulation,
providing a measure of the structural response of the network.

A particularly useful measure of the distinction between bonded
elasticity and steric interactions is the population of strongly
compressed bonds, shown in Figure~\ref{fig5}c. In the absence of an
independent excluded-volume interaction, bonded particles can sample
configurations that are strongly compressed relative to the particle
diameter. The addition of WCA repulsion suppresses such configurations
by introducing a direct energetic penalty for short-range particle
overlap.

The mean bonded separation shown in Figure~\ref{fig5}d provides a
corresponding measure of the average local structure.~\cite{Head2003a,Head2003b,
Storm2005,Broedersz2014} Taken together,
these results demonstrate that the FENE interaction controls the finite 
extensibility and tensile response of the network bonds, whereas the WCA interaction
controls short-range steric organization. The two contributions can
therefore be independently varied without altering the mathematical
form of the bonded interaction.

\subsection{Biological Membrane Networks}

As a final application, we extend the independent FENE formulation
to a mesh-based biological membrane model.~\cite{Marrink2007,
Marrink2013,IzvekovVoth2006,Bidone2019,polley2018snare,polley2025pre,polley2026jcp} Triangulated or
mesh-based coarse-grained representations provide a convenient
framework for describing deformable interfaces in which the
connectivity of the membrane can be separated from additional
mechanical and intermolecular interactions.~\cite{Gompper1994,
Gompper1995,Gompper1997,NoguchiGompper2006,saha2015diffusion,polley2012atomistic,polley2014bilayer} The present formulation
allows finite-extensible elasticity to be assigned directly to the
mesh edges while treating steric interactions independently.

The triangular membrane consists of a regular equilateral mesh in
which each interior vertex is connected to six nearest neighbours.
The same FENE parameters, $K=30$ and $r_0=1.5\,\sigma$, are used for
both interaction models. The Pure FENE membrane contains only the
bonded FENE interaction, whereas the FENE+WCA membrane additionally
includes the short-range excluded-volume interaction.

\begin{figure*}[h!t]
\begin{center}
\includegraphics[width=16.0cm]{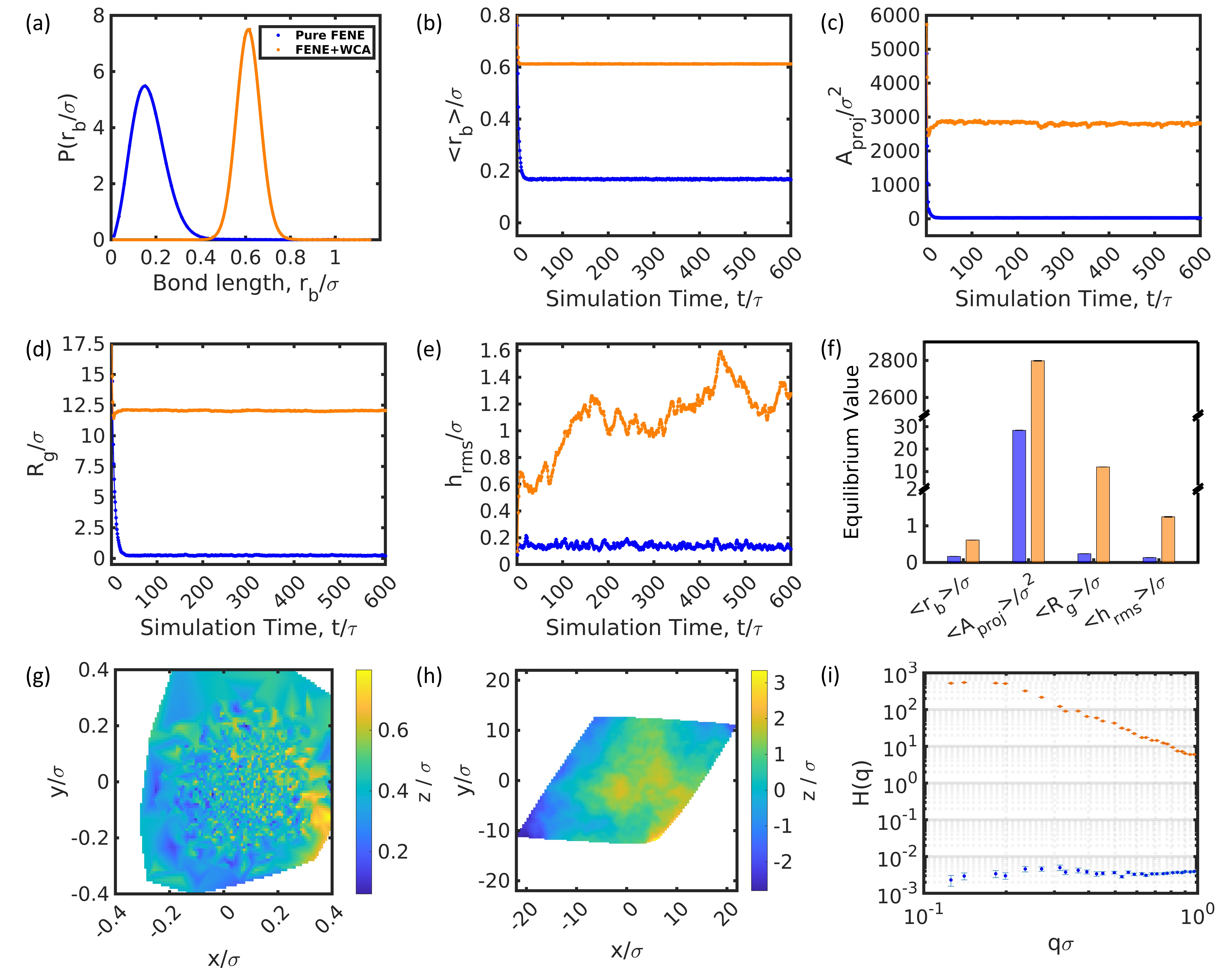}
\caption{
Structural and dynamical characterization of the triangular
biomembrane network with independently controlled bond elasticity and
steric interactions. Pure FENE and FENE+WCA simulations use identical
FENE parameters, $K=30$ and $r_0=1.5\,\sigma$. (a) Representative
membrane configurations. (b) Bond-length distributions,
$P(r_b/\sigma)$. (c) Evolution of the mean bond length during the
simulation. (d) Radius of gyration, $R_g/\sigma$, as a function of
simulation time. (e) Root-mean-square membrane height fluctuation,
$h_{\mathrm{rms}}/\sigma$, as a function of simulation time.
(f) Comparison of the corresponding equilibrium structural
observables. (g,h) Representative spatial distributions of the
membrane height or displacement field for the Pure FENE and
FENE+WCA models, respectively. (i) Height--height fluctuation
spectrum, $H(q)=\langle |h(\mathbf{q})|^2\rangle$, as a function of
the in-plane wave-vector magnitude $q$. The results demonstrate that
finite bond extensibility and short-range steric interactions
constitute distinct physical contributions to the structural and
fluctuation behavior of mesh-based biological membrane models. All
lengths are reported in reduced units of $\sigma$.
}
\label{fig6}
\end{center}
\end{figure*}

Figure~\ref{fig6} summarizes the structural and fluctuation
properties of the triangulated membrane under the two interaction
models. The bond-length distributions demonstrate the local effect
of the additional steric interaction, while the membrane size and
height fluctuations characterize the collective structural response
of the mesh. In particular, the radius of gyration and the
root-mean-square height fluctuation provide measures of membrane
contraction and out-of-plane deformation, respectively.

The Pure FENE membrane undergoes pronounced structural contraction
as the attractive FENE bonds are not opposed by a separate
short-range excluded-volume interaction. In contrast, the addition of
WCA repulsion prevents strong local particle overlap and suppresses
excessive contraction, allowing the membrane to retain a finite
spatial extent.

The membrane application therefore provides a clear demonstration of
the modularity of the independent FENE formulation. The FENE
interaction determines the finite extensibility of the mesh edges,
whereas the WCA interaction provides an independent steric
constraint. Additional membrane interactions, such as bending
elasticity, area or volume constraints, or membrane--protein
interactions, can consequently be introduced without modifying the
bonded FENE potential.~\cite{Helfrich1973,Lipowsky1991,EvansRawicz1990}

To provide a theoretical reference for the measured height--height
fluctuation spectrum, we consider the membrane in the Monge-gauge
approximation, in which the membrane is represented by a single-valued
height field $h(\mathbf r)$ above a reference plane. Within the
small-slope approximation, the quadratic Helfrich Hamiltonian is

\begin{equation}
\mathcal{H}
=
\frac{1}{2}
\int_A d^2r
\left[
\gamma |\nabla h(\mathbf r)|^2
+
\kappa (\nabla^2 h(\mathbf r))^2
\right],
\label{eq:monge_hamiltonian}
\end{equation}

where $\gamma$ is the membrane surface tension and $\kappa$ is the
bending rigidity. Here, $\sigma$ continues to denote the characteristic
particle diameter used as the unit of length throughout the simulations.

Using the Fourier representation

\begin{equation}
h(\mathbf q)
=
\frac{1}{A}
\int_A d^2r\,
h(\mathbf r)
e^{-i\mathbf q\cdot\mathbf r},
\end{equation}

with

\begin{equation}
q=|\mathbf q|
=
\sqrt{q_x^2+q_y^2},
\end{equation}

the Hamiltonian can be written as

\begin{equation}
\mathcal{H}
=
\frac{A}{2}
\sum_{\mathbf q}
\left(
\gamma q^2+\kappa q^4
\right)
|h(\mathbf q)|^2 .
\label{eq:fourier_hamiltonian}
\end{equation}

Application of the equipartition theorem to each independent Fourier
mode gives

\begin{equation}
\frac{A}{2}
\left(
\gamma q^2+\kappa q^4
\right)
\left\langle
|h(\mathbf q)|^2
\right\rangle
=
\frac{k_{\mathrm B}T}{2}.
\end{equation}

Thus, the equilibrium height--height fluctuation spectrum is

\begin{equation}
H(q)
=
\left\langle
|h(\mathbf q)|^2
\right\rangle
=
\frac{k_{\mathrm B}T}
{A\left(
\gamma q^2+\kappa q^4
\right)}.
\label{eq:height_spectrum_theory}
\end{equation}

Accordingly, the tension-dominated regime exhibits

\begin{equation}
H(q)\sim q^{-2},
\qquad
\gamma q^2 \gg \kappa q^4,
\end{equation}

whereas the bending-dominated regime exhibits

\begin{equation}
H(q)\sim q^{-4},
\qquad
\kappa q^4 \gg \gamma q^2.
\end{equation}

The theoretical spectrum therefore provides a continuum reference for
the wave-vector dependence of the membrane height fluctuations
obtained from the simulations. In the present work, $H(q)$ is used
primarily as a measure of out-of-plane membrane fluctuations, while
the independent effects of the bonded and steric interactions are
assessed through the comparison of the Pure FENE and FENE+WCA
systems.

The corresponding height--height fluctuation spectrum, $H(q)$, is
shown in Figure~\ref{fig6}i. The comparison between Pure FENE and
FENE+WCA therefore provides information about how the independent
steric interaction modifies the amplitude and spatial distribution of
membrane height fluctuations. The theoretical basis for $H(q)$ and
the complementary static structure-factor analysis, $S(q)$, are
provided in the Supplementary Information, together with additional
details of the reciprocal-space analysis and supplementary trajectory
movies (Movies S1--S6).

\begin{figure*}[h!t]
\begin{center}
\includegraphics[width=16.0cm]{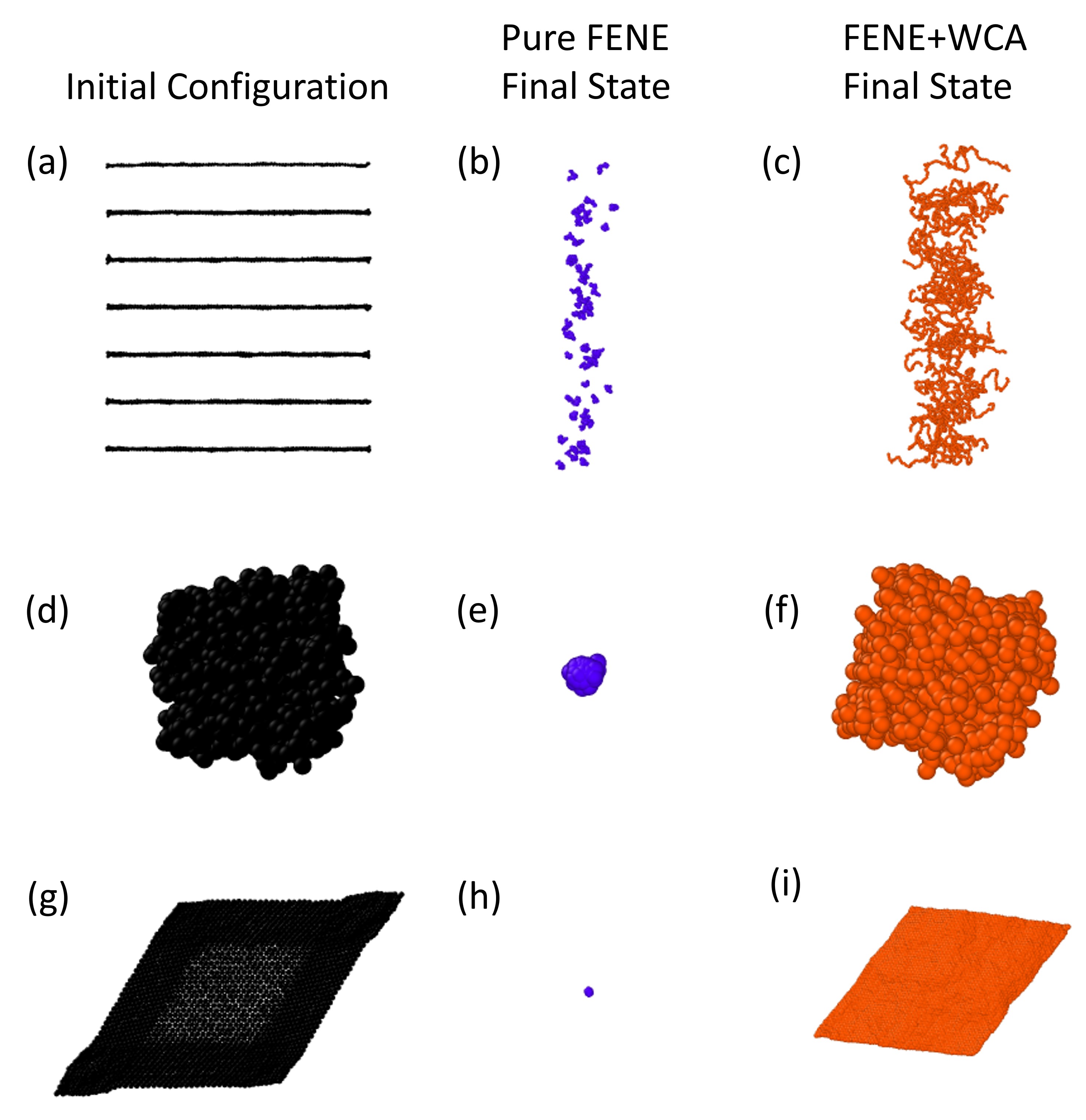}
\caption{
Structural comparison of polymer, colloidal-network, and
triangular-biomembrane systems under Pure FENE and FENE+WCA
interactions. Representative configurations are shown for the
initial state and the final states obtained using Pure FENE and
FENE+WCA interactions. (a--c) Polymer, (d--f) colloidal network, and
(g--i) triangular biomembrane. For each system, the initial
configuration is shown in the left column, while the final
configurations obtained using Pure FENE and FENE+WCA are shown in
the middle and right columns, respectively. The Pure FENE model,
which contains finite-extensible bonded interactions without
short-range excluded-volume stabilization, exhibits pronounced
structural contraction, whereas the addition of WCA repulsion
reduces contraction and preserves a more spatially extended
structure. The corresponding time-resolved structural evolution is provided in
Supplementary Movies S1--S6.
}
\label{fig7}
\end{center}
\end{figure*}

To provide a direct visual comparison of the structural consequences
of the two interaction models across different molecular topologies,
Figure~\ref{fig7} presents representative configurations of the
polymer, colloidal-network, and triangular-biomembrane systems. For
each system, the initial configuration is compared with the
corresponding final configurations obtained using Pure FENE and
FENE+WCA. The Pure FENE systems exhibit pronounced structural
contraction in the absence of short-range excluded-volume
stabilization, whereas the FENE+WCA systems retain more spatially
extended structures. These representative configurations complement
the quantitative analyses presented in Figures~\ref{fig4}--\ref{fig6}.
The corresponding time-resolved trajectory movies are provided in
Supplementary Movies S1--S6.

\section{Conclusion}

We have introduced and validated an independent pure FENE bond
potential for coarse-grained molecular simulations in HOOMD-blue.
The formulation isolates finite-extensible bond elasticity from
nonbonded steric interactions, thereby providing a modular
alternative to conventional FENE+WCA implementations. The analytical
FENE energy and force were derived explicitly and used as exact
benchmarks for validating the numerical implementation. The computed
forces and energies reproduce the analytical expressions to machine
precision over the accessible range of bond extensions.

Comparison with harmonic, FENE+WCA, and harmonic+Lennard--Jones
models further demonstrates that the equilibrium bond distance and
local effective stiffness depend strongly on the interaction terms
included in the model. In particular, Pure FENE describes finite
extensibility without introducing a separate preferred finite bond
length, whereas additional steric or attractive interactions modify
the equilibrium structure and local mechanical response.

Applications to polymer chains, colloidal networks, and mesh-based
biological membrane models demonstrate that the independent FENE
formulation can be applied across substantially different molecular
topologies. In each case, the same finite-extensible bond interaction
can be combined with independently selected nonbonded interactions,
allowing the effects of bond elasticity and steric organization to
be examined separately. The representative configurations further
illustrate that, in the absence of excluded-volume interactions,
Pure FENE bonds can drive pronounced structural contraction, whereas
the addition of WCA repulsion suppresses strong local overlap and
preserves a more spatially extended structure.

The resulting framework is therefore useful for coarse-grained models
in which connectivity and intermolecular interactions arise from
distinct physical mechanisms. In particular, the ability to combine
finite-extensible bonds with independently chosen pair interactions
provides a flexible basis for polymeric materials, colloidal
networks, deformable membranes, and other biomolecular assemblies.

\begin{acknowledgement}

A.P. appreciates the hospitality of generous computing facilities at  SASTRA University, Thanjavur, Tamilnadu. 
A.P. acknowledges the support under Science and Engineering Research Board (SERB), Department of Science and Technology, Government of India [SERB-SRG/2022/001489] and T.R. Rajagopalan research fund, SASTRA University, India.

\end{acknowledgement}

\begin{suppinfo}

Supplementary Information (SI) contains the theoretical height--height
fluctuation spectrum, the static structure-factor analysis of the
biological membrane, and one additional figure ($S1$). Six
supplementary movies (S1--S6) provide time-resolved visualization of
the polymer, colloidal-network, and biological-membrane simulations.

\end{suppinfo}


\providecommand{\latin}[1]{#1}
\makeatletter
\providecommand{\doi}
  {\begingroup\let\do\@makeother\dospecials
  \catcode`\{=1 \catcode`\}=2 \doi@aux}
\providecommand{\doi@aux}[1]{\endgroup\texttt{#1}}
\makeatother
\providecommand*\mcitethebibliography{\thebibliography}
\csname @ifundefined\endcsname{endmcitethebibliography}
  {\let\endmcitethebibliography\endthebibliography}{}

\end{document}